\documentclass[a4paper,12pt]{article}

\usepackage{amscd, amsfonts, amsmath, amssymb, amstext, amsthm, caption, kpfonts, epsfig,fancyhdr, float, graphicx, hyperref, latexsym, mathtools, multicol, multirow,tikz,algorithm,algpseudocode,chngcntr}
\usepackage{subcaption}
\usepackage{authblk}
\usepackage[utf8]{inputenc}
\usepackage{bm}
\usepackage{enumerate}
\makeatletter
\usepackage[margin=2.25cm]{geometry}
\usepackage{color}

\graphicspath{{Figures-Color/}}

\newcommand{\FF}{\mathcal{F}} % PAT: Guessing this one -- need to check
\DeclareMathOperator*{\esssup}{esssup}

\title{Quickest detection in practice in presence of seasonality: An illustration with call center data}
\usepackage{authblk}
\makeatletter
\renewcommand\AB@authnote[1]{\rlap{\textsuperscript{\normalfont#1}}}

\makeatother

\author[(1)]{Patrick J.\ Laub}
\author[(2)]{Nicole El Karoui}
\author[(1)]{St\'ephane Loisel}
\author[(1)]{Yahia Salhi}

\affil[(1)]{{\small Univ Lyon, Université Lyon 1, ISFA, LSAF EA2429}}
\affil[(2)]{{\small Sorbonne Universit\'e, LPSM}}

\date{\today}
\begin{document}

\maketitle

\begin{abstract}

	In this chapter, we explain how quickest detection algorithms can be useful for risk management in presence of seasonality. We investigate the problem of detecting fast enough cases when a call center will need extra staff in a near future with a high probability.
	We illustrate our findings on real data provided by a French insurer. We also discuss the relevance of the CUSUM algorithm and of some machine-learning type competitor for this applied problem.

\end{abstract}
\section{Introduction}
This chapter explores a monitoring scheme to account for possible future breakpoints in the intensity of some Cox-like counting processes in presence of seasonality, and its applicability to assist with call center staffing, thanks to real world data from a call center of a French insurer subsidiary. The practical motivation of this work is the following: is it possible to detect a spike in calls fast enough to mobilize additional staff members and answer the calls as smoothly as possible? El Karoui et al.\ \cite{elkaroui2020a, elkaroui2020b} explained how to use the so-called CUSUM (cumulated sums) algorithm to detect an increase in the claim intensity or in the death intensity for some claim or death process that may feature self-exciting or partially self-inhibiting properties and discuss its implementation in line with Enterprise Risk Management and risk appetite frameworks of the organization. However, those applications were designed to raise an alarm within weeks or months. In the present context, it is necessary to sound the alarm a few minutes or hours after the event, otherwise it is too late to react. The two situations may be comparable or not depending on the average number of events by unit of time. Besides, call center data presents seasonality, and the situation may switch back to normal after a while, which was not the case in El Karoui et al.\ \cite{elkaroui2020a, elkaroui2020b}. Moreover, there might be some time dependence in the present type of dataset, as a problem in the phone network or Internet could breed mass call postponing to the next half day. Therefore, it is interesting to study the practical applicability of the robust detection method and to compare it with machine learning-type competitors. \\

Numerous detection approaches are available in the literature, see \cite{page1954continuous, shiryaev1963optimum, fellouris2012decentralized, siegmund1985sequential} for the discrete-time case and \cite{kailath1998detection} for a detailed review. In the continuous-time we refer to \cite{moustakides2004optimality} and the references therein. As we do not hold enough data on the occurrences of changes and no priors are available, we follow the formulation introduced by Lorden \cite{lorden1971procedures} and investigated for counting processes in \cite{el2017minimax}. We consider that the change occurs at some unknown but deterministic time. Moreover, we assume that the pre-change parameter is an input of the procedure, which may stem from experts' opinions or may be the critical level that would need an intervention from the risk manager. \\

The CUSUM algorithm, initially proposed by \cite{page1954continuous}, is generally used to sequentially detect changes in the distribution of random processes when imposing a Lorden-style constraint on the false alarms. The optimality of the CUSUM algorithm has been established for different cases, both on discrete and continuous time. For example, for independent and identically distributed variables with known distribution before and after the change, \cite{lorden1971procedures} shows that the CUSUM statistics corresponds to the optimal detection strategy. Similarly, \cite{ritov1990decision} use a Bayesian property of the CUSUM statistics and provide an alternative proof of optimality in Lorden's sense. Later on, \cite{moustakides1998quickest} have extended the optimality results to a special class of dependent processes. Optimal CUSUM procedures were proposed by \cite{poor1998quickest} for exponentially penalized detection delays. In continuous time, the optimality of CUSUM has been established for Brownian motion with constant drift by \cite{beibel1996note}, using the Bayesian setting of \cite{ritov1990decision}, which also yielded optimality in Lorden's sense, and by \cite{shiryaev1996minimax}. Recently, an extension of the optimality for the CUSUM test is established by \cite{moustakides2004optimality} to detect changes in It\^o processes when the expected delay is replaced with the corresponding relative entropy in the criterion. Finally, \cite{fellouris2013} establish the optimality of the CUSUM algorithm with respect to a modified version of Lorden's criterion for arbitrary processes with continuous paths and apply this general result to the special case of fractional diffusion–type processes. \\

The case of general doubly stochastic Poisson processes was considered in \cite{el2017minimax}, which follows broadly the framework of \cite{lorden1971procedures} adapted to the case of point processes with a given time-dependent intensity. In the same vein as \cite{shiryaev1996minimax}, \cite{beibel1996note}, \cite{moustakides2004optimality} and \cite{fellouris2013}, it is shown that the CUSUM stopping time is still optimal when one wants to minimize the average number of events until detection (instead of the average delay before detection). Aside from the theoretical results on the optimality, this framework is also interesting from the practical point of view due to the large use of the Cox's regression model in the life actuarial fields especially for experienced mortality forecasting. Therefore, in this chapter we use the framework of El Karoui et al.\ \cite{el2017minimax} to investigate the surveillance procedure for counting processes whose intensity exhibits seasonality. We introduce the quickest detection problem of such seasonal count processes and discuss the main challenges when implementing the CUSUM surveillance procedure. \\

The remainder is organized as follows. In \autoref{sec:detection}, we explain in detail how to implement the CUSUM algorithm in presence of seasonality. \autoref{sec:case_study} is dedicated to the case study. In \autoref{sec:comparison}, we compare the pros and the cons of the CUSUM strategy with the ones of alternative strategies, using modified versions of the real world CUSUM. The code used in this chapter is available at \url{https://www.github.com/Pat-Laub/SeasonalCUSUM}.

\section{Detection in counting processes with seasonality} \label{sec:detection}

\begin{figure}[t]
	\centering
	\begin{tikzpicture}
		\draw (0, 0) node {\includegraphics[width=.8\textwidth]{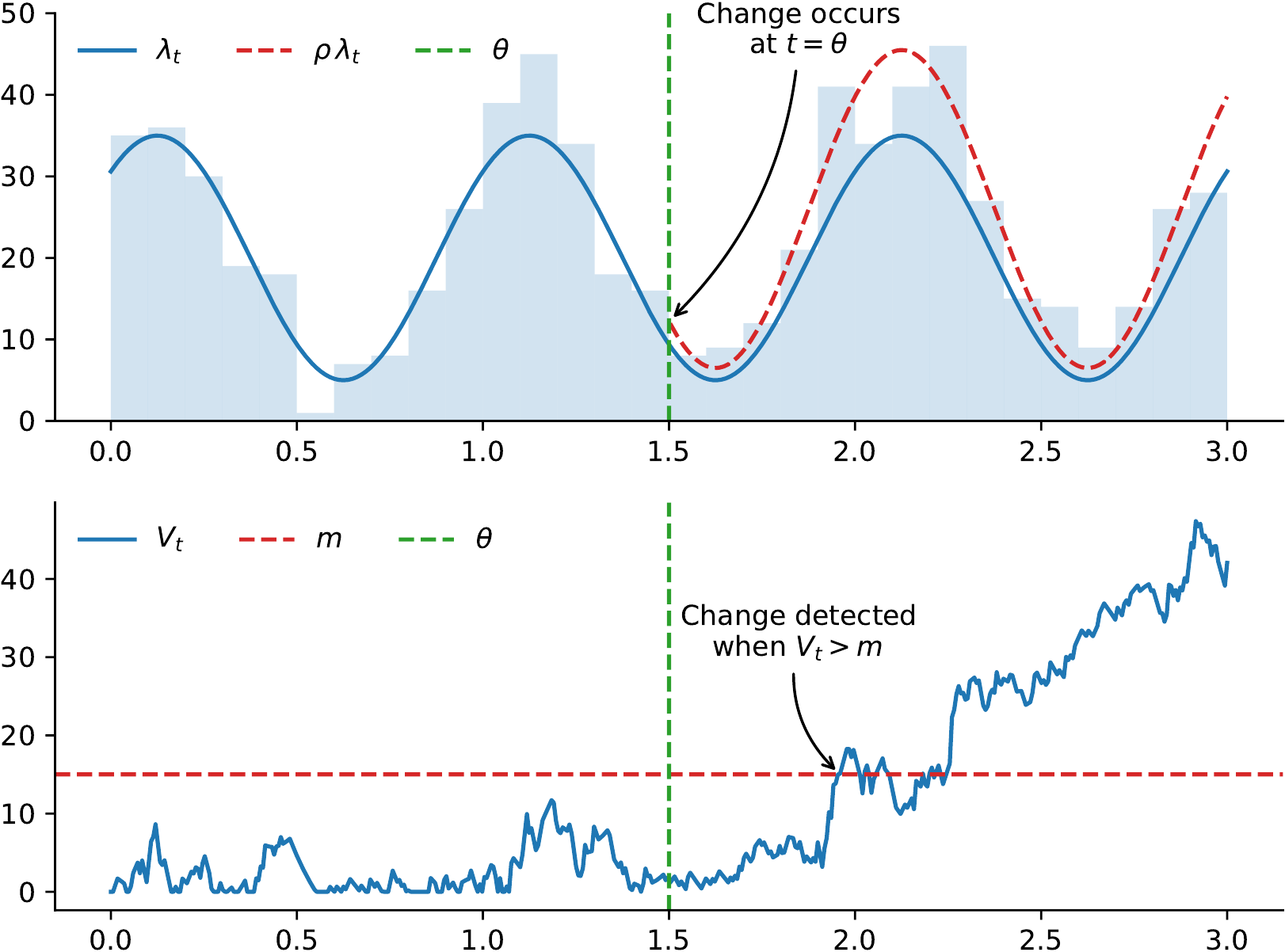}};
		\draw (-7, 4.9) node {(a)};
		\draw (-7, -0.1) node {(b)};
	\end{tikzpicture}
	\caption{(a) Example of seasonal intensity (solid line) with simulated records and a change-point at time $\theta=1.5$ with $30\%$ increase of the intensity (dashed line). (b) Sample path of the CUSUM process $V$.}%with continuous (black) and discretely observed observations (red).}
	\label{fig:intensity}
\end{figure}

In most applications, data exhibits strong seasonality. A wide variety of business, economic, medical and actuarial processes can be represented by seasonal time series models. An example can be seen in contagious diseases data such as dengue, influenza-like illnesses and malaria, among others, that have a strong seasonal pattern in most regions of the world. In actuarial applications, claims records show a seasonal behavior so that there are variations between the summer and the winter seasons. Also, more often, frequency changes are driven by day-of-the-week and other usual influences, see \cite{verrall2016understanding}. In our case, we are interested in the surveillance of telephone call centers, where an accurate understanding of call arrivals is crucial for the efficient staffing of such centers. The data at our disposal reports the number of customer calls at a given periodicity. Here, we work on records with half-hourly and daily spaced intervals. For such data, empirical research has suggested that it is appropriate to model the arrival process as an inhomogeneous Poisson process, see \cite{weinberg2007bayesian}. However, the intensity exhibits a seasonality pattern. For instance, this could have the form in \hyperref[fig:intensity]{\autoref*{fig:intensity}a}. In this section, we introduce the quickest detection problem of such seasonal count processes and discuss the main challenges when implementing the CUSUM surveillance procedure.

\subsection{Understanding the seasonal patterns} \label{sec:seasonality}

Although the inhomogeneous Poisson arrival specification for some datasets is straightforward, it is important to understand the underlying theoretic justifications and exhibit the main specific features of its intensity. To this end, we will consider the dataset used in the following sections which gathers records of customer call arrivals of a French call center. Even though the discussion hereafter focuses on call centers specific datasets, the underlying model and technical development could apply to many other settings, such as epidemics, insurance claims and many more. Throughout the sequel, the dataset we will use consists of aggregate number of calls over a given period. During the weekdays, the call center receives calls from 7:30 to 18:30. In this organization, the Saturday opening hours are shorter than the weekday operating hours (7:30 to 12:30). For both periods, we have at our disposal the number of calls each half-hour. Even if the call-by-call arrival process is not observed, the Poisson assumptions are equivalent to saying that individual calls are independent of each other, conditional on the arrival rate. As noted earlier, there are many cases where this assumption makes sense, at least to a good approximation, see \cite{el2017minimax} for a discussion. \\

\begin{figure}[t]
	\centering
	\begin{subfigure}{0.45\textwidth}
		\centering

		\includegraphics[width=\textwidth]{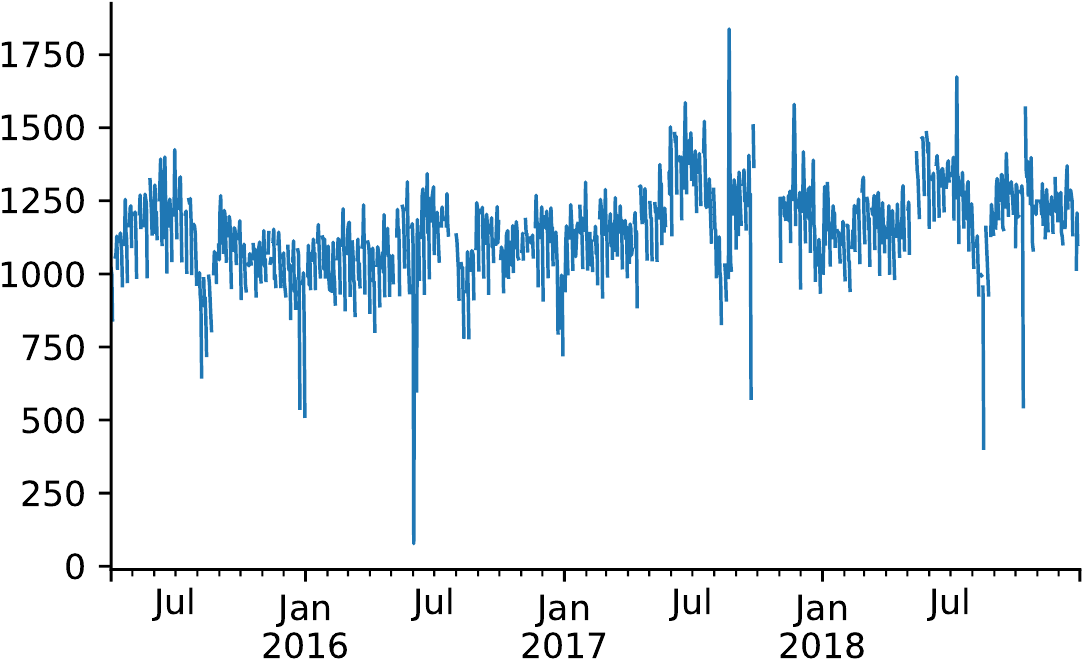}
		\caption{Number of calls each weekday over the period ranging from April 2015 to January 2019.}
		\label{subfig:years}
	\end{subfigure}\hfill
	\begin{subfigure}{0.45\textwidth}
		\centering
		\includegraphics[width=\textwidth]{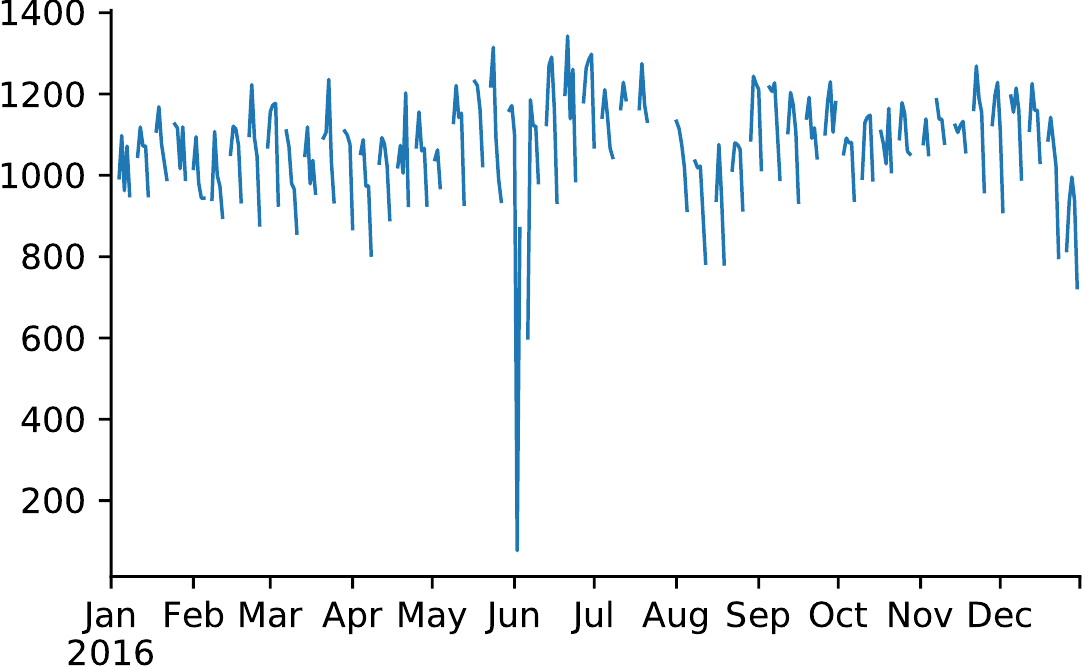}
		\caption{Number of calls each weekday over the year 2016.}
		\label{subfig:first_year}
	\end{subfigure}

	\vspace{0.25cm}

	\begin{subfigure}{0.45\textwidth}
		\centering
		\includegraphics[width=\textwidth]{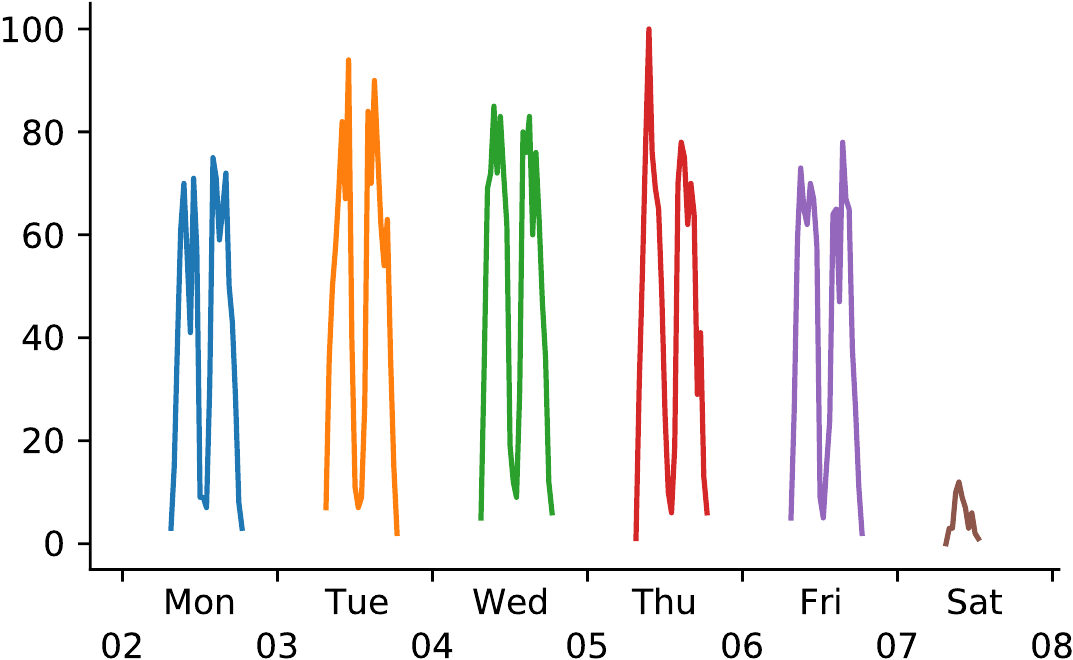}
		\caption{Number of calls during the week ranging from Monday 2nd to Saturday 7th of January 2017.}
		\label{subfig:week}
	\end{subfigure}\hfill
	\begin{subfigure}{0.45\textwidth}
		\centering
		\includegraphics[width=\textwidth]{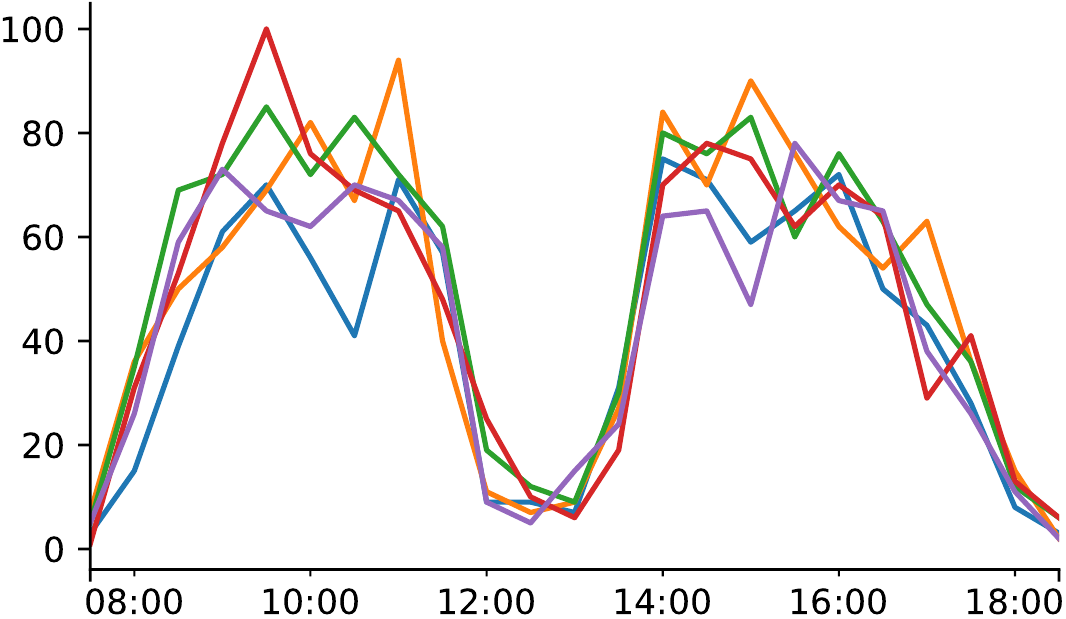}
		\caption{Number of calls during each day from Monday 2nd to Friday 6th of January 2017.}
		\label{subfig:day}
	\end{subfigure}
	\caption{Seasonality in call arrivals: (a) yearly, (b) monthly, (c) daily and (d) hourly.}
	\label{fig:data}
\end{figure}

In the case of call centers datasets, the arrival rate is varying over time and typically exhibit intra-day, daily, monthly and yearly seasonality. \autoref{fig:data} illustrates the arrival patterns that are typically observed. In \autoref{subfig:years} and \autoref{subfig:first_year}, we show the yearly evolution of the calls. We see that there are monthly fluctuations in the data, and that there is perhaps a small increase in call volume over the years in the considered period. When we observe the monthly data over 2016, we observe that there is an intra-year seasonality.
% This brings up the first key component in modeling such datasets, which is the trend. This is observed when looking over longer time horizons.
On the other hand, when observing the data in shorter weekly (\autoref{subfig:week}) or daily (\autoref{subfig:day}) intervals, we see that the temporal structure of the arrival rate has a two-way character, since there is day-to-day variation and also intra-day variation. All weekdays have similar daily profiles, with two major daily peaks for call arrivals, see \autoref{subfig:week}. To recap, the arrival rate is time-varying, a feature which is not accounted for in a homogeneous Poisson arrival process. It is, therefore, natural to consider an inhomogeneous extension and let the intensity depend on time. Henceforth, the arrival rate is denoted $\lambda=(\lambda_t)_{t\geq 0}$ and will incorporate both strong within-day patterns and day-to-day dynamics. %We note that, in the literature, it is commonly assumed that the intensity $\lambda$ is constant over short time periods. In this paper, we will not make such an assumption so that the intensity could be modeled as a deterministic or stochastic process.

\subsection{Monitoring call arrivals and change-point detection problem}\label{subsec:detection}

Suppose one continuously monitors the customers telephone calls process which is initially ``in control''. This means that the customers calls are in line with the regular and normal arrival rates, which of course accommodate the aforementioned seasonality. At some future time point this process may go ``out of control'' and it is then desirable to react. The objective is to detect any departure from this controlled environment. In other words, the goal is the detection of future workload peaks. %, which is of paramount importance for efficient staffing of such centers.
Suppose that we sequentially observe the process $N=(N_t)_{t\geq 0}$ counting the telephone calls and let $\mathbb{F}=(\mathcal{F}_t)_{t\geq 0}$ be the observable information, i.e.\ in the $\sigma$-algebra generated by the process $N$. We will concentrate our attention on the model in which the observed counting process $N$ has an intensity that undertakes a sudden and proportional change at time $\theta\in [0, \infty]$. When the process is \textit{in control} its intensity is given by $\lambda_t$. When the change occurs, the intensity becomes $\rho \lambda_t$, where $\rho$ is a predefined and constant parameter. In our case,
% as mentioned above,
we are interested in an increase of the call arrival rate and thus assume that $\rho > 1$. More formally, we have the following form of the intensity for $t\geq 0$
\begin{equation}\label{eq:intensity}
	\lambda_t 1_{\{t<\theta\}} + \rho\lambda_t 1_{\{t\geq \theta\}} \,.
\end{equation}
In \hyperref[fig:intensity]{\autoref*{fig:intensity}a}, we show a sample path of the intensity with a change-point at time $\theta=1.5$. At this time, the calls arrival rate suddenly increases by $30\%$ (dashed line) compared to the normal and expected pattern (solid line). \\

In order to locate the change-point $\theta$, we adopt the same framework as in \cite{el2017minimax}. Thus, the unknown change-point $\theta$ is assumed to be deterministic. This is mainly motivated by the lack of any prior on the behavior of the change-points in view of past observations of $N$. The goal is then to sound an alarm as soon as the change happens. Formally,
given that the process is observed sequentially and assuming that the proportional change size $\rho$ is known, the objective is to find an $\FF$-stopping time $\tau$ which has the smallest detection delay. To design such a procedure, we will need two types of performance measures: one being a measure of the delay between the time a change occurs and the time it is detected, and the other being a measure of the frequency of false alarms. Following \cite{el2017minimax}, we would like to find $\tau$, if it exists, with minimum detection delay, measured through the worst case of detection delay. In order to understand this framework, we will consider the probability measure $\mathbb{P}_{\theta}$ under which the process $N$ has the intensity described in \autoref{eq:intensity} for a given $\theta$. In the case $\theta=\infty$ the probability measure $\mathbb{P}_\infty$ corresponds to the case where the process is \textit{in control}, i.e.\ the change will never occur. Similarly, $\mathbb{P}_0$ corresponds to \textit{out of control} situation, i.e.\ the change occurs at time at $\theta=0$. We further agree on some notation we will use throughout the chapter and denote by $\mathbb{E}_\theta$ the expectation with respect to $\mathbb{P}_\theta$. Using this notation, the optimal detection time is constructed as the one minimizing the following delay criterion
$$\sup_{\theta \geq 0} \esssup \mathbb{E}_{\theta}\left[(N_\tau-N_\theta)^+ \,|\, \mathcal{F}_t\right].$$
Here, the worst-case detection delay is adopted as a measure of the detection lag, conditioned on the observations before the change time $\theta$. Such a criterion is a generalization of the Lorden's criterion \cite{lorden1971procedures}, taking into account the inhomogeneity of the underlying Poisson process. This is, in fact, shown to be more convenient for optimally designing an optimal procedure based on the so-called cumulative sums (CUSUM) method introduced by \cite{page1954continuous}. Indeed, as proved by \cite{el2017minimax}, the CUSUM process can be used to characterize the optimal stopping rule. To this end, the desire to make the above quantity small, i.e.\ quick detection, should be balanced with a constraint on the rate of false alarms measured by the average run length given by $\mathbb{E}_\infty[N_\tau]$. Hence, the optimal stopping times $\tau$ satisfies the false alarm constraint
$$\mathbb{E}_\infty[N_\tau] = \pi,$$
where $\pi > 0$ is the predefined level of false alarm.

In order to introduce the CUSUM-based optimal stopping scheme, given the aforementioned performance measures, we need to define the \textit{log sequential probability ratio} process, denoted $U=(U_t)_{t\geq 0}$, between the reference probability $\mathbb{P}_\infty$ (null assumption of no change, i.e.\ $\theta=\infty$) and the alternative assumption $\mathbb{P}_0$ (H1 assumption with an immediate change, i.e.\ $\theta=0$). The process $U$ can be written for $t\geq 0$ as
\begin{equation}\label{eq:process_U}
	U_t=N_t-\beta(\rho) \Lambda_t,
\end{equation}
where $\beta(\rho)=(\rho-1)/\log\rho$ and $\Lambda_t=\int_{0}^{t}\lambda_s \,\mathrm{d}s$. The CUSUM process $V=(V_t)_{t\geq 0}$ is defined as the reflected version of $U$. More precisely, the process $V$ can be written, for each $t\geq 0$, as
\begin{equation}\label{eq:process_V}
	V_t=U_t-\inf_{0 \le s \le t} U_s.
\end{equation}
This process allows us to construct the optimal stopping rule $\tau^V_m$, called the CUSUM stopping time, in the sense of \cite{el2017minimax}. Indeed, the CUSUM stopping rule is defined as the first passage of the process $V$ to level $m$, i.e.\ $\tau^V_m=\inf\{t\geq 0, V_t\geq m\}$. To ensure the optimality, the level $m$ is chosen to fulfill the false alarm constraint. In other words, we should fix this level such that $\mathbb{E}_\infty[N_{\tau^V_m}]=\pi$.

\subsection{Practical application of the CUSUM method}\label{subsec:applicability}

%
%\begin{figure}[t]
%	\centering
%	\begin{tikzpicture}
%		\draw (0, 0) node {\includegraphics[width=.8\textwidth]{granularity.pdf}};
%		\draw (-6.8, 5) node {(a)};
%		\draw (-6.8, 0) node {(b)};
%	\end{tikzpicture}
%	\caption{The approximate CUSUM processes (a) $U_t$ and (b) $V_t$ when the data is only observed on a grid $t_i = i \Delta t$.}
%	\label{fig:granularity}
%\end{figure}

As discussed in \autoref{sec:seasonality}, the dataset of interest exhibits two main features: a strong seasonality and possibly a weak persistent trend. The scheme discussed above aims to detect a proportional increase of the intensity, thus, potential changes in the trend are not handled in this chapter. An optimal alarm for a change in the intensity of magnitude $\rho$, under the false alarm constraint, requires the construction of the process $V$, which can be simply represented based on the knowledge of the call's occurrences and the assumption on the intensity. In fact, as showed by \cite{el2017minimax}, the CUSUM process $V$ obeys the following pathwise differential equation:
\begin{equation*}
	\mathrm{d} V_t = \mathrm{d}N_t-\beta(\rho) \, 1_{(0,\infty)}(V_t) \, \lambda_t \, \mathrm{d}t.
\end{equation*}
This means that the process $V$ has the same behavior as $U$ in \autoref{eq:process_U}, with jumps of size $1$ at the times of each call arrival. In between jumps, $V$, decreases at rate $-\beta(\rho)\lambda_t$ which is cut off when $V = 0$. In other words, the reflection only intervenes when $U$ hits $0$, at which $V$ is equal to $0$ and remains so until the next jump of $U$. To see this particular pattern, we depict a sample path of the process $V$ in \hyperref[fig:intensity]{\autoref*{fig:intensity}b}. Here, the intensity $\lambda_t$ is the one shown in \hyperref[fig:intensity]{\autoref*{fig:intensity}a} in the solid line. The intensity changes at time $\theta=1.5$ by the factor $\rho=1.3$. On the set $\{t<\theta\}$, the CUSUM process $V$ remains close to the lower level $0$ as the change has not yet occurred, and the process departs from $0$ after the change. \\

We should note that this sample is built based on the continuously observed count process $N$ and the full knowledge of the exact intensity $\lambda$. In practice, the process $V$ cannot always be constructed with perfect accuracy as the arrival times of the calls may be aggregated; this is frequently the case in many practical applications. Thus, one must be able to construct the process $V$ from that information, knowing that the process $N$ is only observed at discrete times $t_i$ at intervals ranging from half an hour to a day in the particular case of interest. With such incomplete information, one cannot determine with certainty for how long the CUSUM process has been staying at zero and whether the CUSUM process has crossed the detection barrier $m$. Therefore, in the present study, we focus on the time-granular dataset and it would be much more difficult to use another dataset with incomplete information (see El Karoui et al.\ \cite{elkaroui2020a, elkaroui2020b} for more details). \\

The detection procedure is closely tied to the knowledge of the arrival rate $\lambda$. We, thus, should be able to anticipate the intra-day intensity taking into account the stylized facts discussed in the previous section. For instance, in order to handle seasonal effects, we can use a part of the historical data that we are currently in as a basis to predict the expected count for next period. This will give the anticipated behavior of the call arrivals at time points with similar conditions. However, we should note that any usual seasonal shifts that were not anticipated in the intensity can interfere with the sensitivity of the CUSUM detection rule. This can be either amplified or obscured during the detection procedure. The importance of the intensity estimation is thus of paramount importance. In the next sections, we will emphasize the estimation procedure and investigate the impact on the detection procedure.

\section{Case study: Call center data}\label{sec:case_study}

\subsection{Preliminary analysis of call center dataset}

\begin{figure}[t]
	\centering
	\begin{subfigure}{\textwidth}
		\centering
		\begin{tikzpicture}
			\draw (0, 0) node {\includegraphics[width=.8\textwidth]{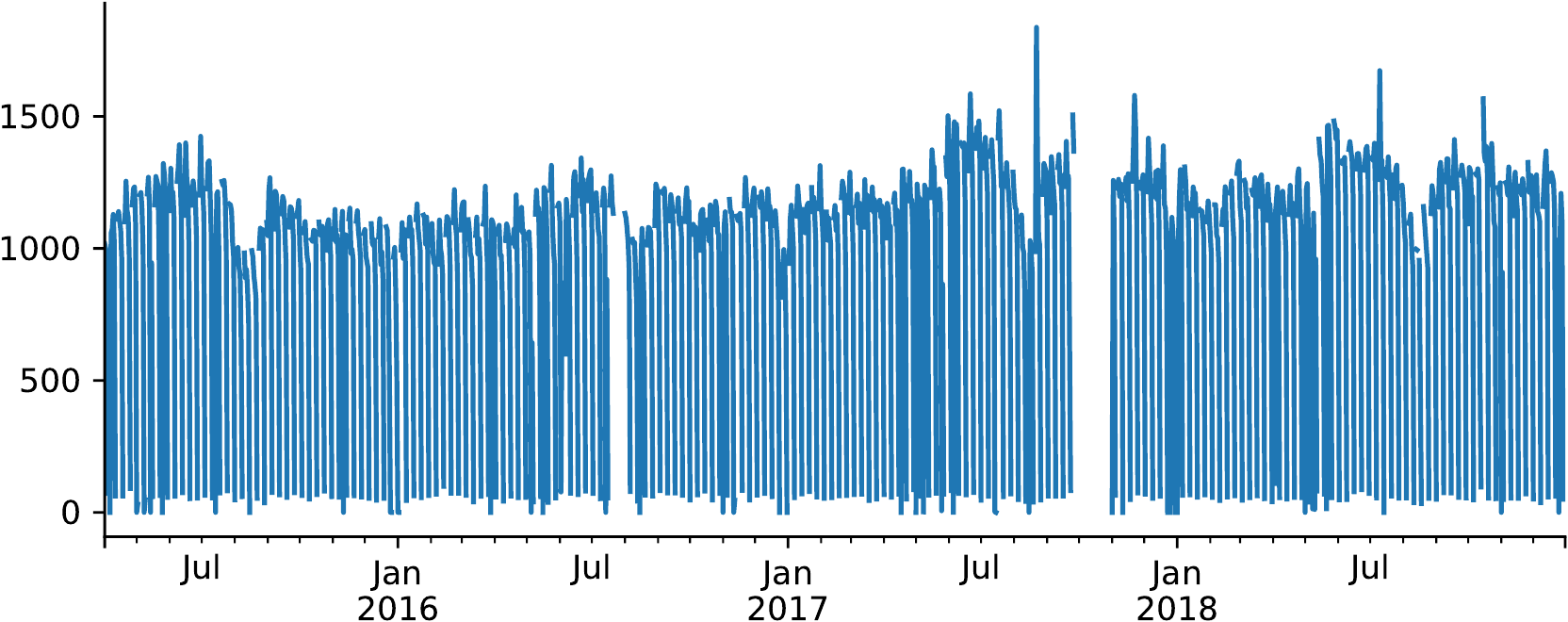}};
			\draw (-7, 2.5) node {(a)};
		\end{tikzpicture}
	\end{subfigure}
	\begin{subfigure}{\textwidth}
		\centering
		\vspace{-1em}
		\begin{tikzpicture}
			\draw (0, 0) node {\includegraphics[width=.8\textwidth]{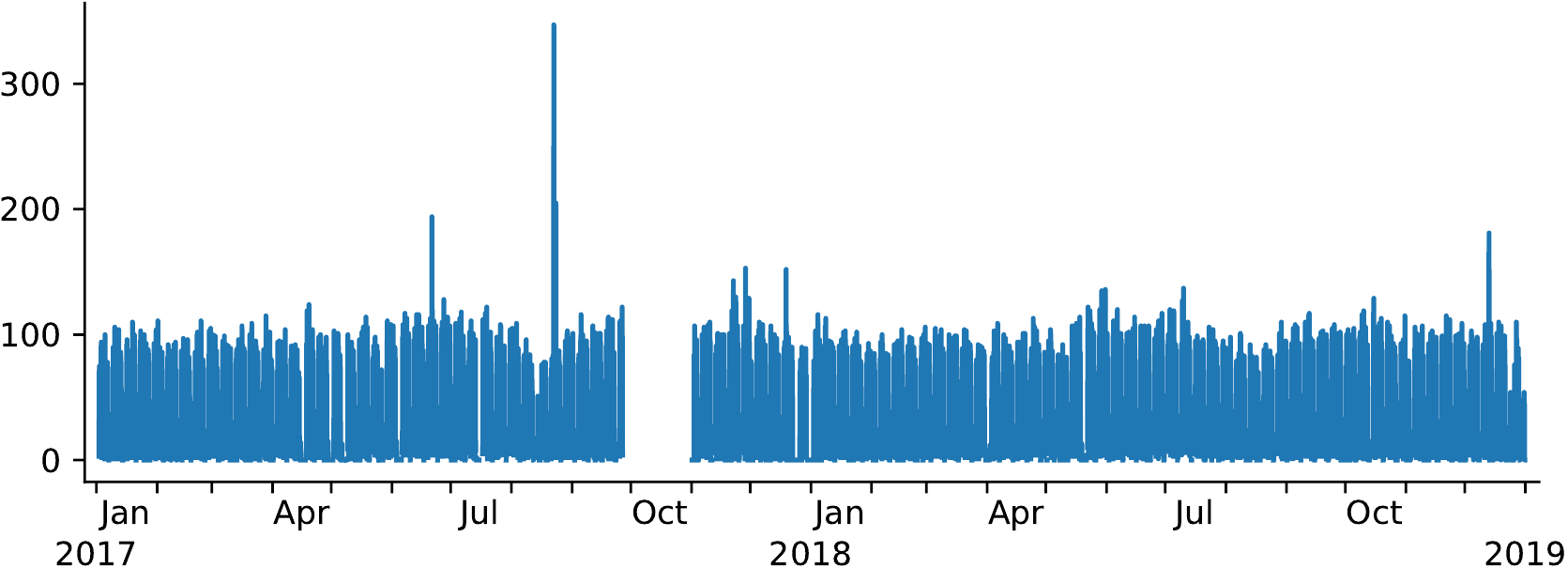}};
			\draw (-7, 2.5) node {(b)};
		\end{tikzpicture}
	\end{subfigure}
	\caption{The number of calls received by the call center aggregated by (a) daily and (b) half-hourly intervals.}
	\label{fig:overview}
\end{figure}

We consider a dataset provided by a French insurer, detailing the number of calls arriving to their European call centers. The number of calls aggregated daily was available from 2015-04-06 until 2018-12-31, and the number of calls aggregated by half-hour blocks (between 7:30 until 18:30) was available between 2017-01-02 and 2018-12-31. We will refer to these as the daily dataset and the half-hourly dataset, respectively. The two datasets are shown in \autoref{fig:overview}. As discussed in the previous section, looking at this figure, we can draw three main conclusions: i) the arrival process exhibits strong seasonality, ii) there appears to be no significant long-term trend, and iii) there are large sections of missing data (e.g.\ the month of October in 2017). \\

The call center is open from morning to evening on Mondays to Fridays, and during the morning on Saturdays; a representative week of the data is shown in \autoref{subfig:week}. There is a large disparity in call arrivals between the weekdays and the Saturdays. From \hyperref[fig:overview]{\autoref*{fig:overview}a}, we see that the most days see between 1000 and 1500 calls, but on each Saturday the graph plummets towards zero calls to reflect the paucity of weekend calls. In comparison \autoref{subfig:years}, which shows the same data with Saturdays removed, is considerably more stable. Simply knowing whether a particular day is a weekday or a Saturday explains nearly 97\% of the variance of the number of calls arriving on that day. \\

\begin{figure}[t]
	\centering
	\includegraphics[width=0.9\textwidth]{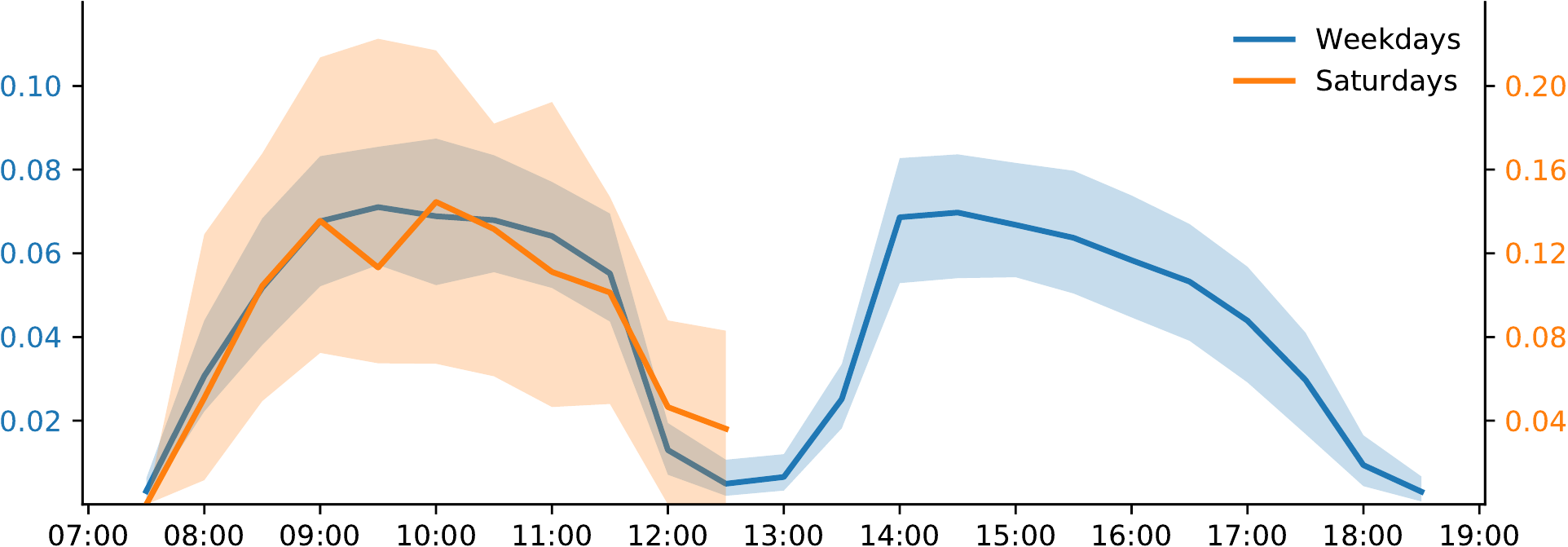}
	\caption{The median fraction of each day's calls which falls in each half-hour time slot. As the call centers only accept calls on Saturdays in the mornings, we plot the weekdays separately to the Saturdays. The shaded regions fill between the 5\% and 95\% quantiles of the different fractions.}
	\label{fig:frac-calls-halfhour}
\end{figure}

The next obvious seasonality is the intra-daily patterns. The representative week shown in \autoref{subfig:day} highlights the typical behavior, where most calls arrive either between 9:00--12:00 or between 14:00--17:00, with few calls arriving in the intervening period. This pattern (mostly) holds for Saturdays also. To accommodate these anomalous Saturdays, we look at the number of calls arriving in each half-hour as a fraction of that day's total calls. \autoref{fig:frac-calls-halfhour} shows the median fraction of daily calls arriving in each half-hour, where weekdays and Saturdays are treated separately. We can conclude that the half-hourly distribution of calls is basically the same for weekdays and Saturdays. For example, about 7\% of a weekday's calls arrive between 9:00 and 9:30, and about 14\% of a Saturday's calls arrive in the same period; as Saturdays are only open for half the day, the two patterns agree. \\

Finally, there is a lot of missing data in the two datasets. In one sense, the half-hour dataset is missing the first couple of years of data as it starts in 2017 yet the daily dataset begins in 2015. As noted above, the month of October in 2017 is missing in both datasets. We will take this as a natural split in the data and call the period before this missing month the training set, and the period after the test set. In the next section, we will use the training set to fit a model to predict the number of calls to arrive in each half-hour period. This prediction model will be the $\lambda_t$ in our CUSUM experiments. We will run the CUSUM algorithm on the test set, to try to find change-points in the data which is new, i.e.\ which wasn't used to fit the $\lambda_t$ model.

\subsection{Fitting a model for the number of calls in each half hour}

From the preliminary analysis, we can tell that the number of arrivals is strongly influenced by the half-hour period of the day, by the day of the week (with Saturday being a clear outlier), by the month of the year, and possibly by a long-term trend. As such, we will fit a sequence of models to handle these various levels of seasonality. We split the data into a training and a test set, estimate the expected number of arrivals on the training set, and consider CUSUM detection on the test set. \\

We fitted a sequence of generalized linear models (GLMs) for the number of calls arriving each day. The canonical logarithmic link function is used. The following factors were considered:
\begin{itemize}
	\item whether or not the day is a weekday or a weekend day (boolean),
	\item the number of days since the start of the daily dataset (integer),
	\item the month (categorical),
	\item the day of the week (categorical),
	\item whether or not the day immediately follows a holiday (boolean).
\end{itemize}
\begin{table}[t]
	\caption{A comparison of five generalized linear models to predict the number of daily call arrivals with a logarithmic link function. Each row represents a model with the selected independent factors, where smaller BIC values are better.}
	\label{tbl:glm-comparison}
	\centering
	\begin{tabular}{c|c|c|c|c|c}
		\begin{tabular}[c]{@{}c@{}}Mon. to\\ Fri.\end{tabular} & \begin{tabular}[c]{@{}c@{}}Days since\\ 2015-04-01\end{tabular} & Month        & \begin{tabular}[c]{@{}c@{}}Day of \\ Week\end{tabular} & \begin{tabular}[c]{@{}c@{}}Day after\\ Holiday\end{tabular} & BIC  \\ \hline
		$\checkmark$               &                            &              &                            &                            & 9340 \\
		$\checkmark$               & $\checkmark$               &              &                            &                            & 8303 \\
		$\checkmark$               & $\checkmark$               & $\checkmark$ &                            &                            & 5640 \\
		                           & $\checkmark$               & $\checkmark$ & $\checkmark$               &                            & 2781 \\
		                           & $\checkmark$               & $\checkmark$ & $\checkmark$               & $\checkmark$               & 2668
	\end{tabular}
\end{table}

The daily counts are considered here because our daily dataset covers a longer time period (it starts about two years before our half-hourly dataset); the training part of the half-hourly dataset is shorter than one year, so it is impossible to fit any monthly or yearly seasonality just using this data. \autoref{tbl:glm-comparison} shows the Bayesian information criterion (BIC) for five candidate models which used different combinations of these factors.  \\

\begin{figure}[t]
	\centering
	\includegraphics[width=0.8\textwidth]{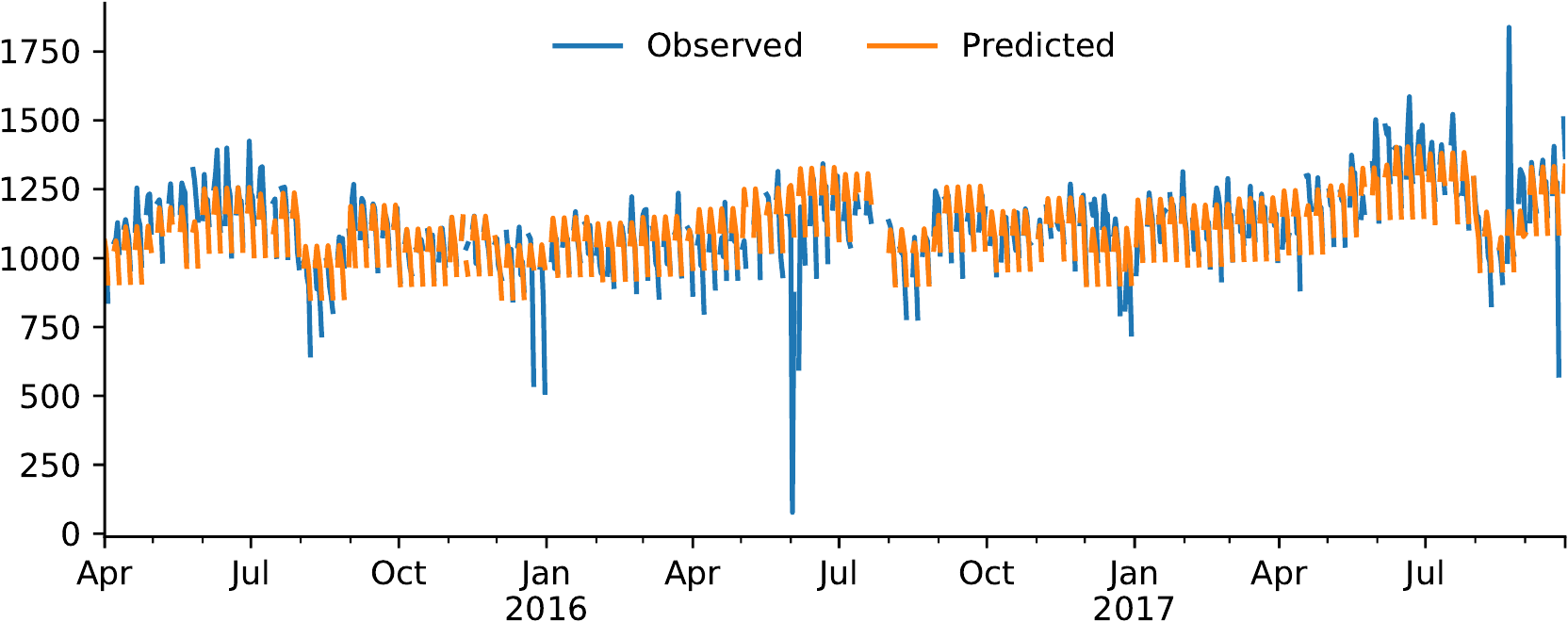} \\
	\caption{The number of calls each day in the training part of the daily dataset and the GLM's fitted number of calls. For readability, only weekdays are plotted (i.e.\ Saturdays are excluded).}
	\label{fig:glm-train}
\end{figure}

\begin{figure}[t]
	\centering
	\begin{subfigure}{0.4\textwidth}
		\caption{Training set}
		\vspace{-0.5em}
		\includegraphics[width=\textwidth]{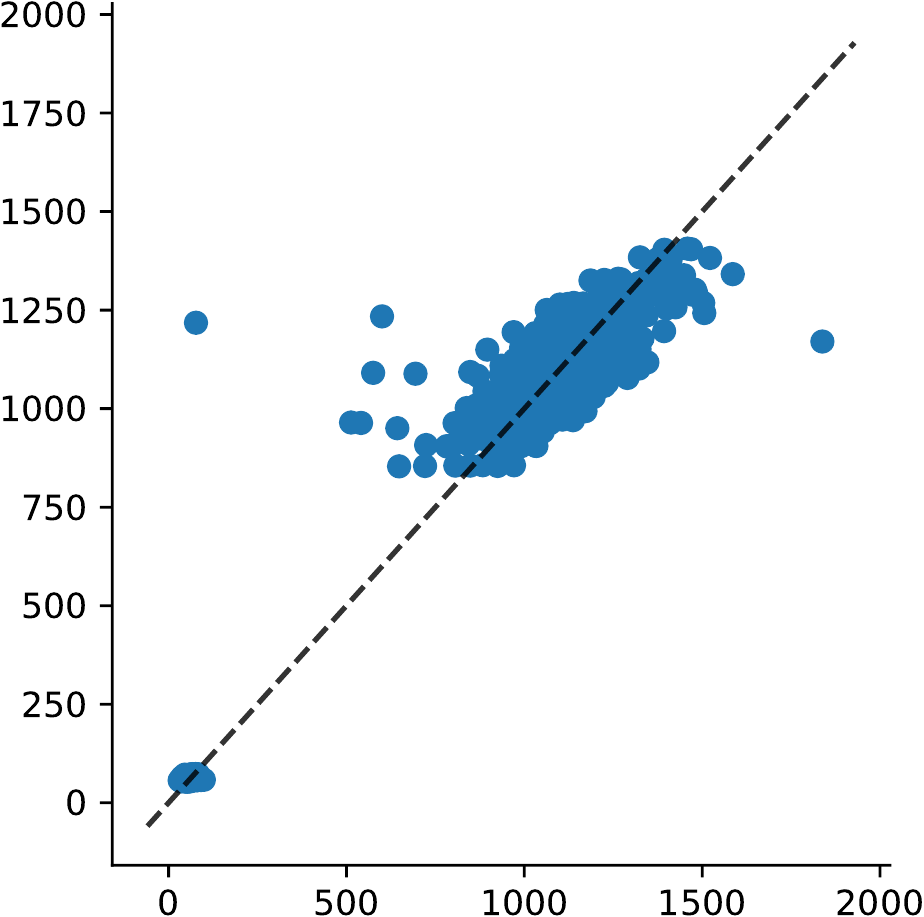}
	\end{subfigure}
	\begin{subfigure}{0.4\textwidth}
		\caption{Test set}
		\vspace{-0.5em}
		\includegraphics[width=\textwidth]{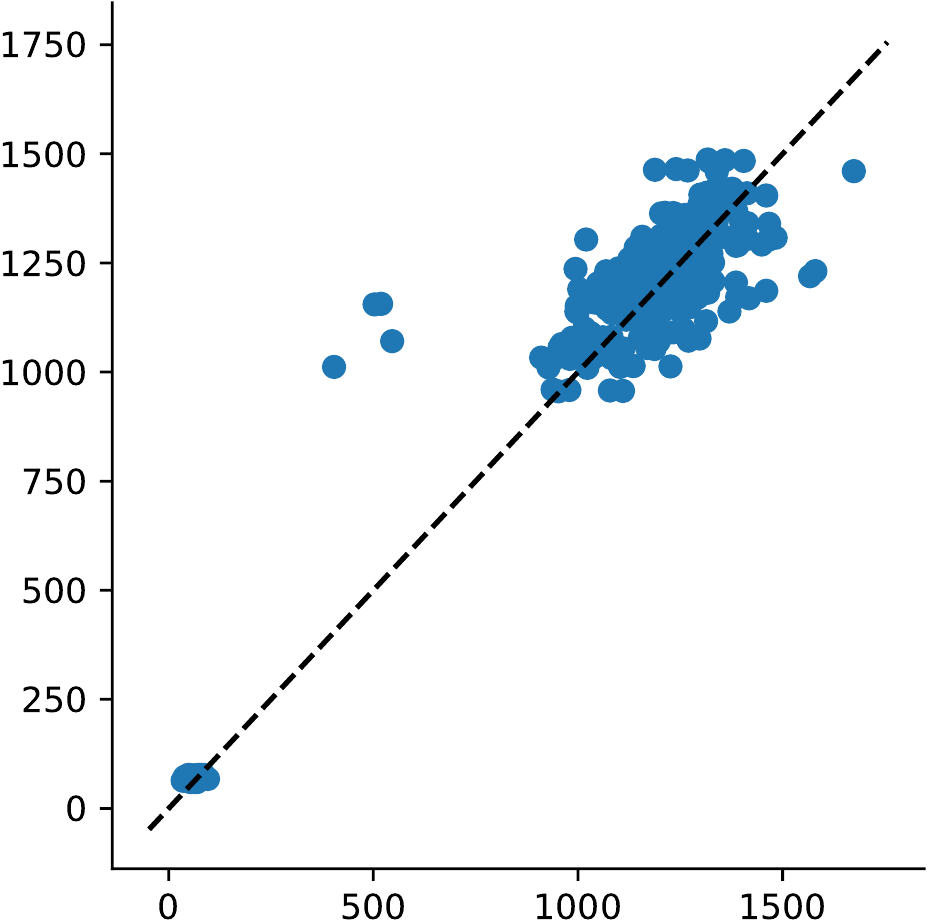}
	\end{subfigure}
	\caption{A scatterplot of the observed number of calls ($x$-axis) against the number predicted by the GLM ($y$-axis). A good fit will have the points fall on or close to the diagonal.}
	\label{fig:glm-scatterplots}
\end{figure}

The model in the final row of \autoref{tbl:glm-comparison}, which is the model with the smallest BIC, was selected. \autoref{fig:glm-train} compared this selected GLM's `predictions' to the observed data in the training set. We see that it fits the training data quite well, exhibiting the main features of the data we observed in the preliminary analysis, without appearing to overfit the data. \autoref{fig:glm-scatterplots} compares the GLM predictions against the observations as a scatterplot, where the training data is shown separately to the test data. The similarity of the two scatterplots reinforces our conclusion that the GLM is not overfitting the training data. Note that the choice of a GLM in this step is not required for the CUSUM method. Any prediction mechanism would have sufficed. We fitted a gradient boosted decision tree (XGBoost) as a comparison \cite{XGBoost}, but it underperformed the GLM so it was discarded.
\\

The GLM is fitted to predict the number of calls that will arrive in a day, though we wish to run the CUSUM method with a time resolution of half-hour periods. Thus, we need to allocate these predicted daily calls to the half-hour time slots. In \autoref{fig:frac-calls-halfhour} of our preliminary analysis, we noted that the distribution of each day's calls across each half-hour time slot is quite stable (noting that Saturday requires special treatment). \\

\begin{figure}[t]
	\centering
	\includegraphics[width=0.9\textwidth]{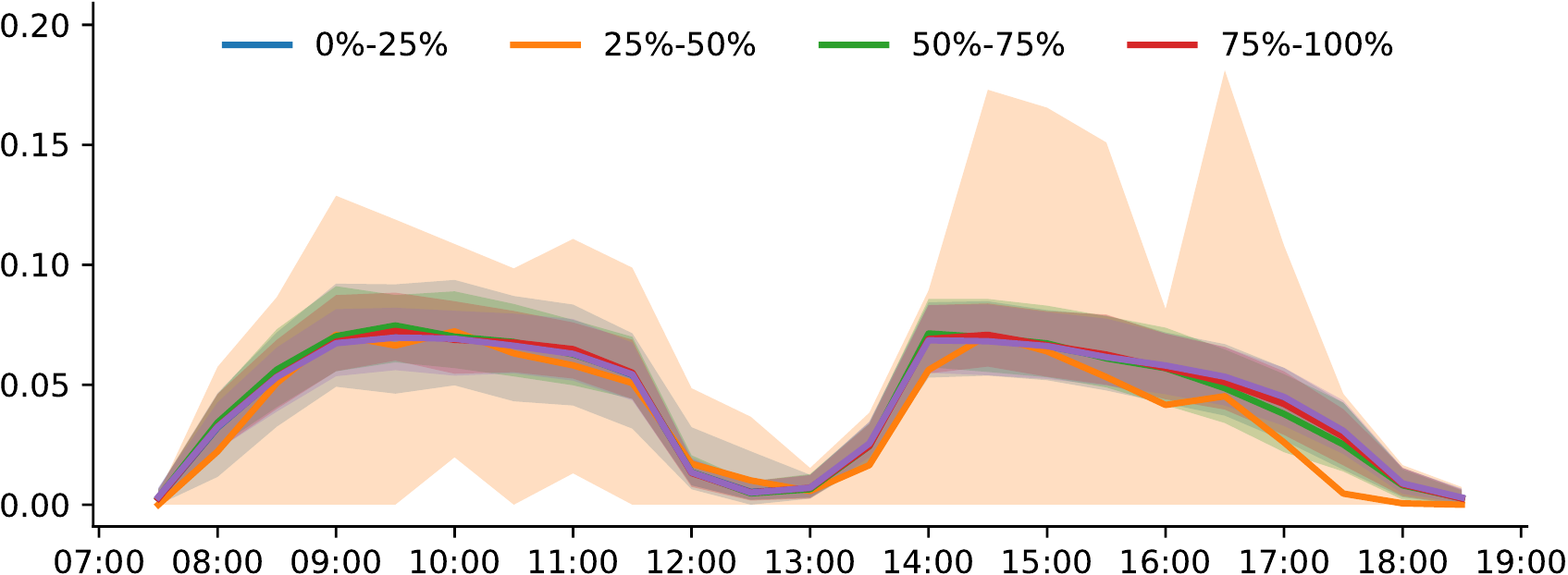}
	\caption{The median fraction of each day's calls which falls in each half-hour time slot, where the days are grouped by similar levels of busyness. The shaded regions fill between the 5\% and 95\% quantiles of the groups.
	}
	\label{fig:frac-calls-by-busyness}
\end{figure}

If we assume that the fraction of each day's calls which arrives in the different half-hour time slots is independent of the total number of calls each day, then we can simply take the daily predictions from the GLM and multiply them by the historical fraction which arrived in each time slot. This results in a simple method to construct $\lambda_t$, but this assumption of independence could easily be incorrect --- e.g.\ it might be that days with more calls may have a spike of calls in the morning, as compared to quieter days where the calls arrive more evenly through the day. To assess this, we split the days into groups of busyness, and then plot the fraction of calls which arrives during each half-hour period in each group. Specifically, days are split into the quartiles of the observed daily arrival counts. The result is in \autoref{fig:frac-calls-by-busyness}. Although some groups (like the `moderately quiet' days whose count falls between the 25\% and 50\% quantiles) have higher variance than others, the average values are quite stable, so our independence assumption appears justified.

\subsection{CUSUM detection ignoring the seasonality}

\begin{figure}[t]
	\centering

	\begin{subfigure}{\textwidth}
		\centering
		\begin{tikzpicture}
			\draw (0, 0) node {\includegraphics[width=.9\textwidth]{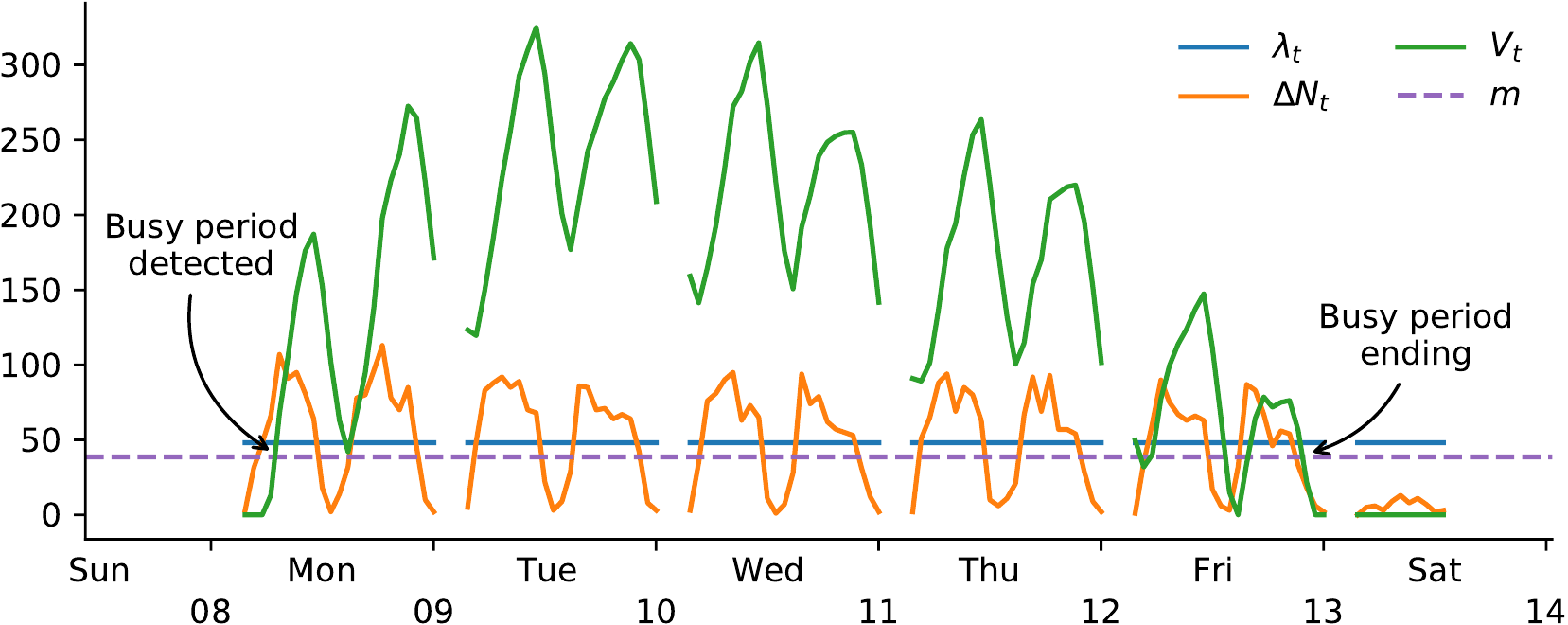}};
			\draw (-7.1, 3.1) node {(a)};
		\end{tikzpicture}
	\end{subfigure}
	\begin{subfigure}{\textwidth}
		\centering
		\vspace{-0.5em}
		\begin{tikzpicture}
			\draw (0, 0) node {\includegraphics[width=.9\textwidth]{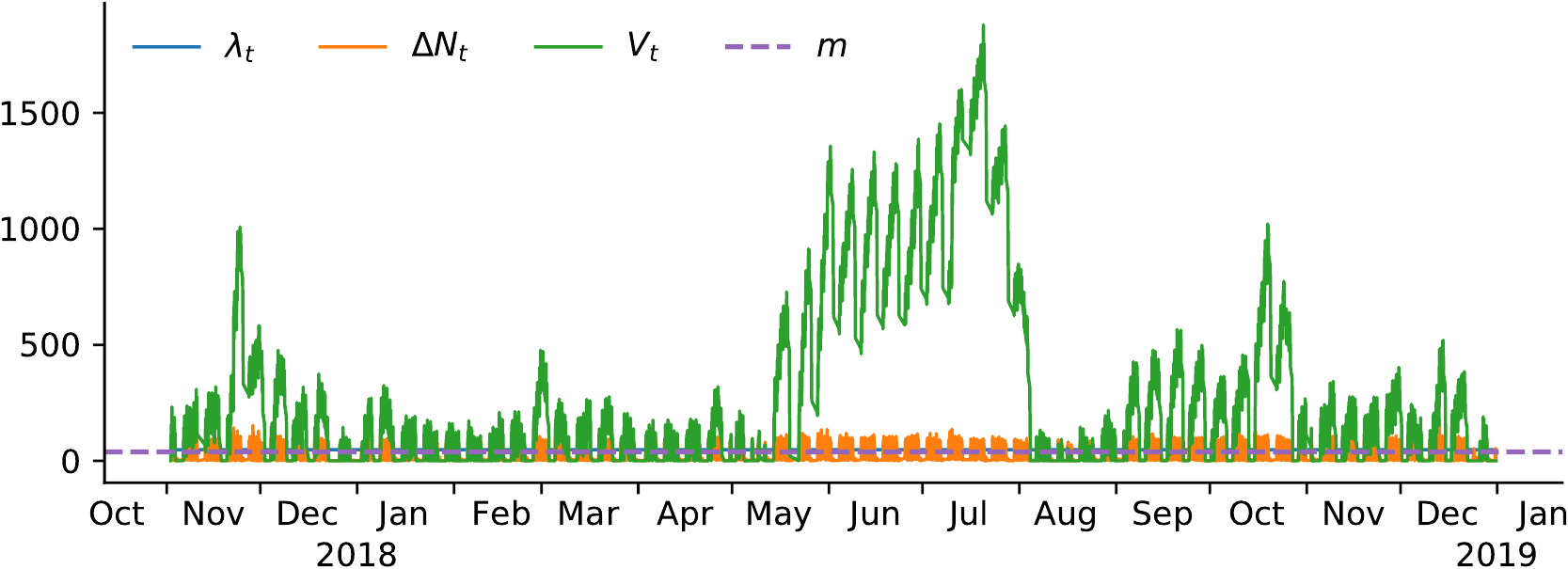}};
			\draw (-7.05, 2.6) node {(b)};
		\end{tikzpicture}
	\end{subfigure}
	\caption{The results of CUSUM with a naive $\lambda_t \equiv \lambda$ over (a) the week starting 2018-01-08 and (b) the whole test period. The number of calls arriving in each half-hour time slot is denoted $\Delta N_t$.}
	\label{fig:naive-cusum}
\end{figure}

In the next section we will run the CUSUM algorithm using the GLM-based $\lambda_t$, but first let's see how CUSUM behaves when the seasonality in the data is totally ignored and we use the most simplistic $\lambda_t$. We run CUSUM with $\lambda_t \equiv \lambda$ where $\lambda$ is the sample average of the number of calls per half hour in the training dataset, and we attempt to detect a multiplicative change of size $\rho = 1.2$. \hyperref[fig:naive-cusum]{\autoref*{fig:naive-cusum}a} shows the CUSUM output for the first complete week of 2018 which is pretty representative. Here, the CUSUM alarm is raised on the Monday, and the alarm is still activated until the Saturday when the warning level plummets to zero. This reflects the unsurprising fact that nearly every weekday is $\ge 20\%$ busier than the long-run average, and that Saturdays are generally much quieter than this average. Thus, a change-point is detected nearly every week. The CUSUM threshold $m = 38.7$ was chosen so that a false-positive alarm should be raised roughly once every year. Considering \hyperref[fig:naive-cusum]{\autoref*{fig:naive-cusum}b} though, the $V_t$ process is consistently very large during the test period. In fact, the $m$ threshold is exceeded for about 80\% of the time! \\

\subsection{CUSUM detection with seasonality}

\begin{figure}[t]
	\centering
	\includegraphics[width=0.9\textwidth]{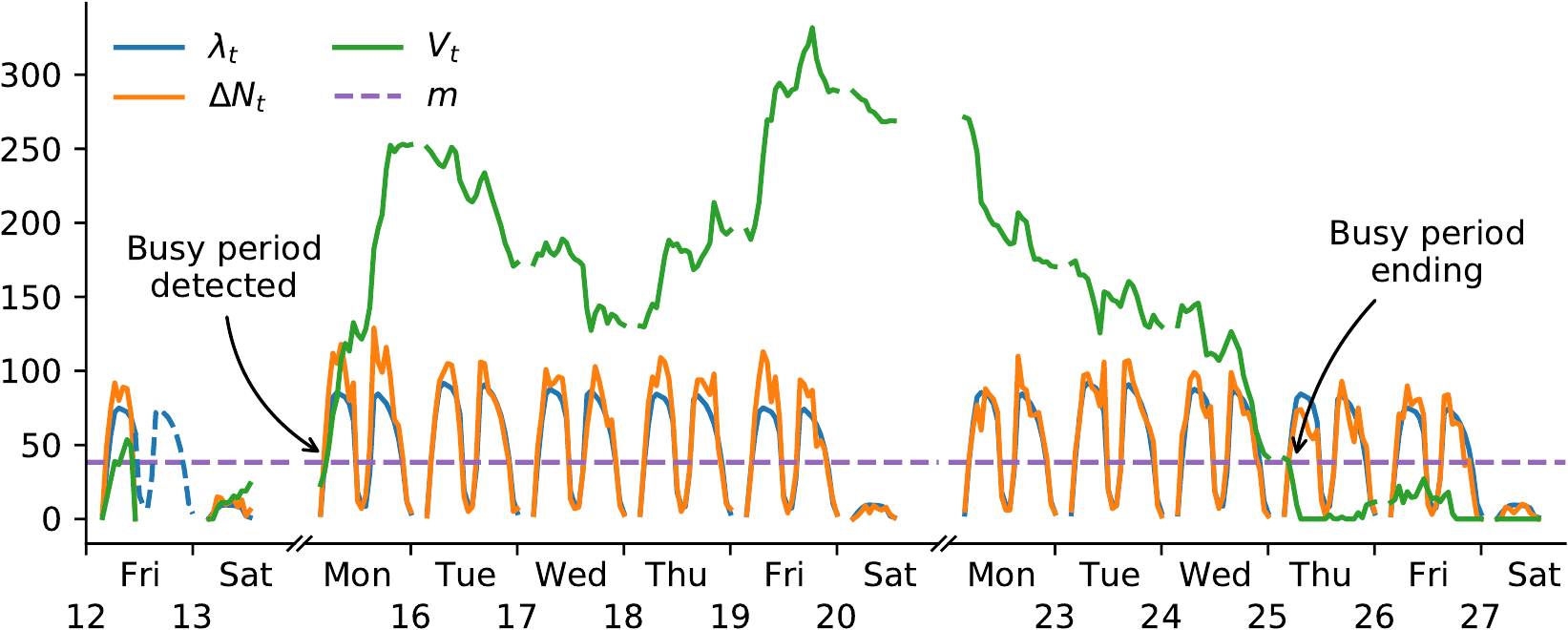}
	\caption{CUSUM algorithm results for a few weeks starting from 2018-10-12. The call center closed at noon on the Friday the 12th, which is unusual. In the CUSUM we modified $\lambda_t$ to be zero for that afternoon, though here we plotted the original $\lambda_t$.}
	\label{fig:cusum-case-study-2018-10-12}
\end{figure}

Obviously, the naive model for $\lambda_t$ in the previous section is too simplistic for this highly seasonal dataset. We rerun the CUSUM algorithm when the expected number of arrivals $\lambda_t$ fluctuates based on the GLM scheme described earlier. One successful detection in the CUSUM results occurs around Friday 2018-10-12, which is illustrated in \autoref{fig:cusum-case-study-2018-10-12}. The Friday shows a normal morning, but the call center appears closed in the afternoon. This leads to an increase in calls on the Saturday which raises the warning level. Then the following Monday and Tuesday are busy days, and the warning level stays high until Tuesday afternoon when the activity has returned to normal. \\

\begin{figure}[t]
	\centering
	\includegraphics[width=0.9\textwidth]{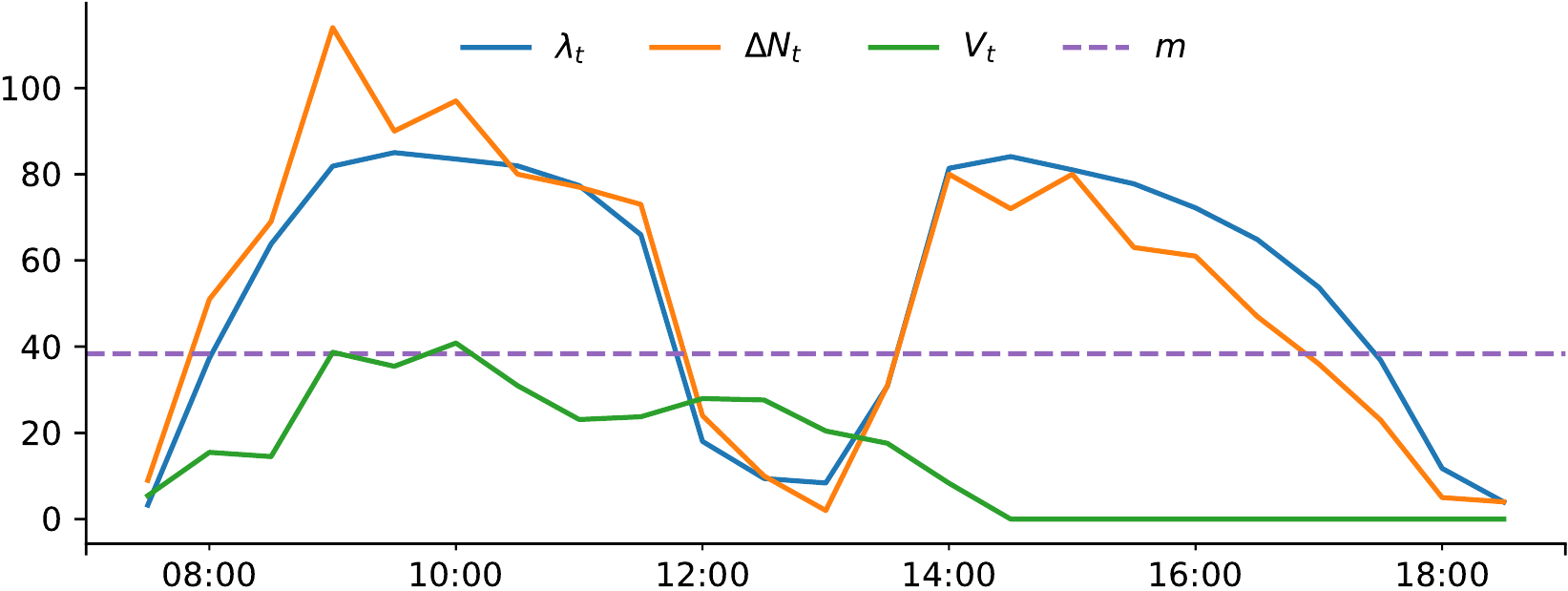}
	\caption{CUSUM algorithm results for the date 2018-06-15.}
	\label{fig:cusum-case-study-2018-06-15}
\end{figure}

The detection shown in \autoref{fig:cusum-case-study-2018-10-12} represents a best-case scenario of the CUSUM algorithm. An alarm is raised swiftly, and so the call center managers could react to the significant surge of calls which occur in the 1--2 weeks following the alarm. Unfortunately, this best-case behavior is quite rare. \autoref{fig:cusum-case-study-2018-06-15} shows a more common and less desirable result. On 2018-06-15 the call center received a moderate surge in calls in the morning, and the CUSUM raises an alarm around 9:00. Yet this surge dissipates as soon as the alarm is raised, so if extra call center operators are asked at short notice to work in the afternoon, then they will have missed this ephemeral spike in calls.

\section{Discussion of CUSUM and of some alternatives}\label{sec:comparison}

The theory guarantees that the CUSUM is optimal (in the generalized Lorden sense) even in the presence of seasonality, as long as the increments of the counting process are independent. Even if El Karoui et al.\ \cite{el2017minimax} have shown that CUSUM remains optimal for a large class of counting processes with path-dependent intensity (for example Hawkes processes), if the time dependence of increments is not taken into account correctly in the intensity, then the CUSUM is suboptimal because the model is incorrect from the beginning. If CUSUM is ``trained'' on the right model incorporating this path-dependency, then it remains optimal. To illustrate this practical issue, we consider two such situations. \\

\begin{figure}[t]
	\centering
	\begin{subfigure}{\textwidth}
		\centering
		\begin{tikzpicture}
			\draw (0, 0) node {\includegraphics[width=.8\textwidth]{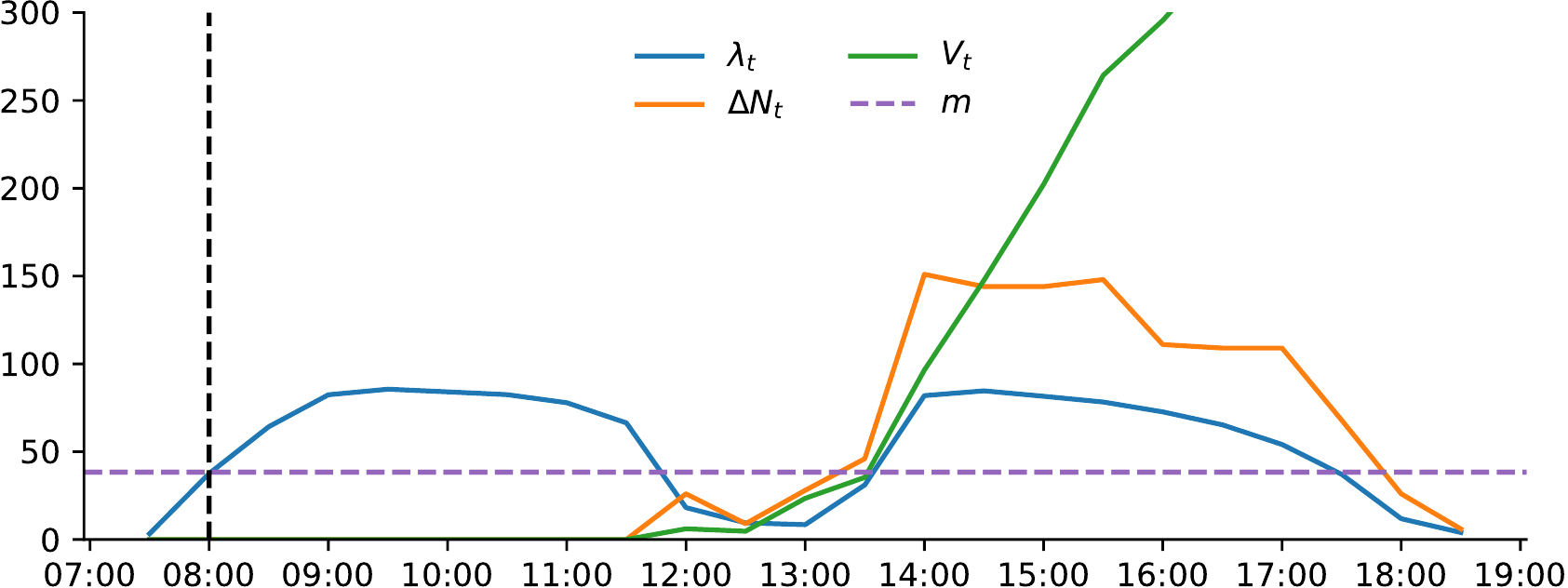}};
			\draw (-7, 2.5) node {(a)};
		\end{tikzpicture}
	\end{subfigure}
	\begin{subfigure}{\textwidth}
		\centering
		\vspace{-1em}
		\begin{tikzpicture}
			\draw (0, 0) node {\includegraphics[width=.8\textwidth]{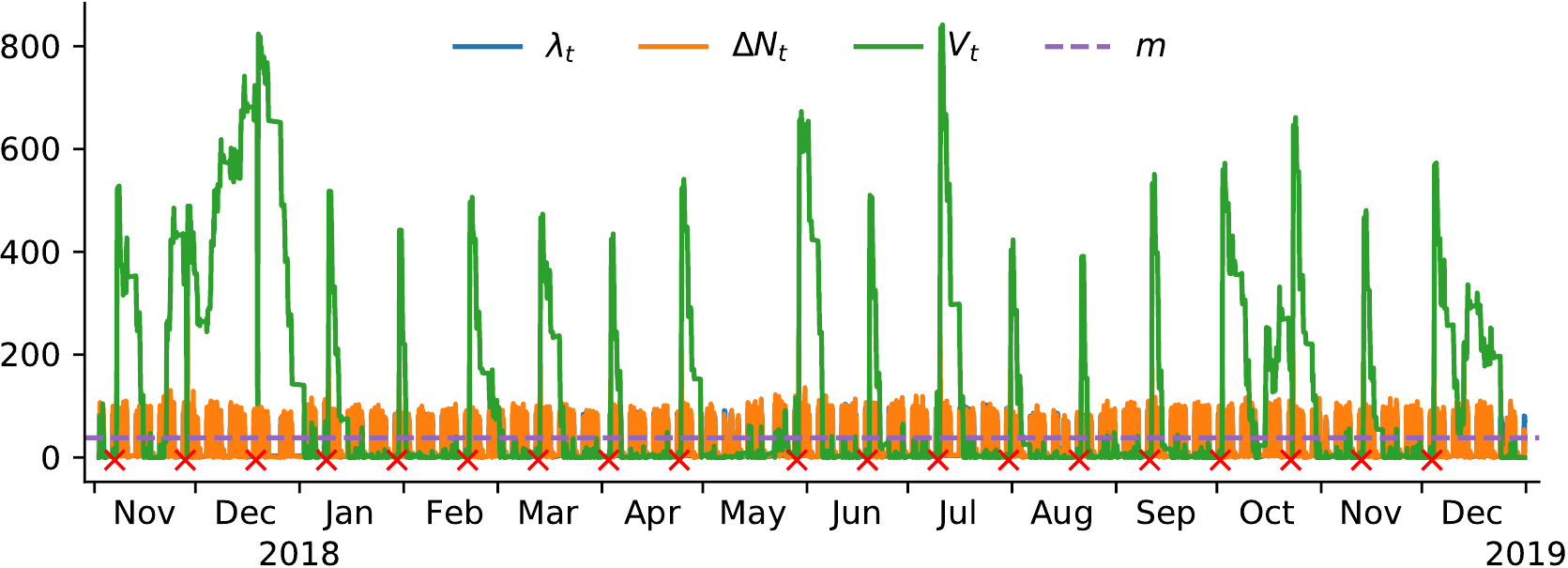}};
			\draw (-7, 2.5) node {(b)};
		\end{tikzpicture}
	\end{subfigure}
	\caption{The CUSUM results over (a) 2018-04-03 and (b) the whole test set, where every third Tuesday (marked with a red cross) has had the morning calls shifted into the afternoon.}
	\label{fig:modified-tuesdays}
\end{figure}

In our context, it could happen that some calls are postponed, due to some failure in the telecommunication system for example. Some data analysis with generalized linear models shows that such an effect is not really present in our dataset. To see how the CUSUM would react if the seasonality was badly estimated, or if some calls could be postponed, we artificially modify our dataset: for the morning of every third Tuesday, calls are postponed to the corresponding afternoon. This represents of course a big change in the seasonality.
\hyperref[fig:modified-tuesdays]{\autoref*{fig:modified-tuesdays}a} shows the breakdown of CUSUM for one such Tuesday, as a change is detected at around 13:30. Looking across \hyperref[fig:modified-tuesdays]{\autoref*{fig:modified-tuesdays}b} we can see that the alarm is sounded on each of these modified Tuesdays. If one assumes that a closed Tuesday morning is always followed by an afternoon surge, then, as seen in \hyperref[fig:modified-tuesdays]{\autoref*{fig:modified-tuesdays}a}, it would be possible to detect at 8:00 (vertical dotted line) the need for extra staff members in the afternoon (instead of 13:30). In a similar manner, a second CUSUM algorithm detecting a decrease in the intensity would do a similar job. \\

\begin{figure}[t]
	\centering
	\begin{subfigure}{\textwidth}
		\centering
		\begin{tikzpicture}
			\draw (0, 0) node {\includegraphics[width=.8\textwidth]{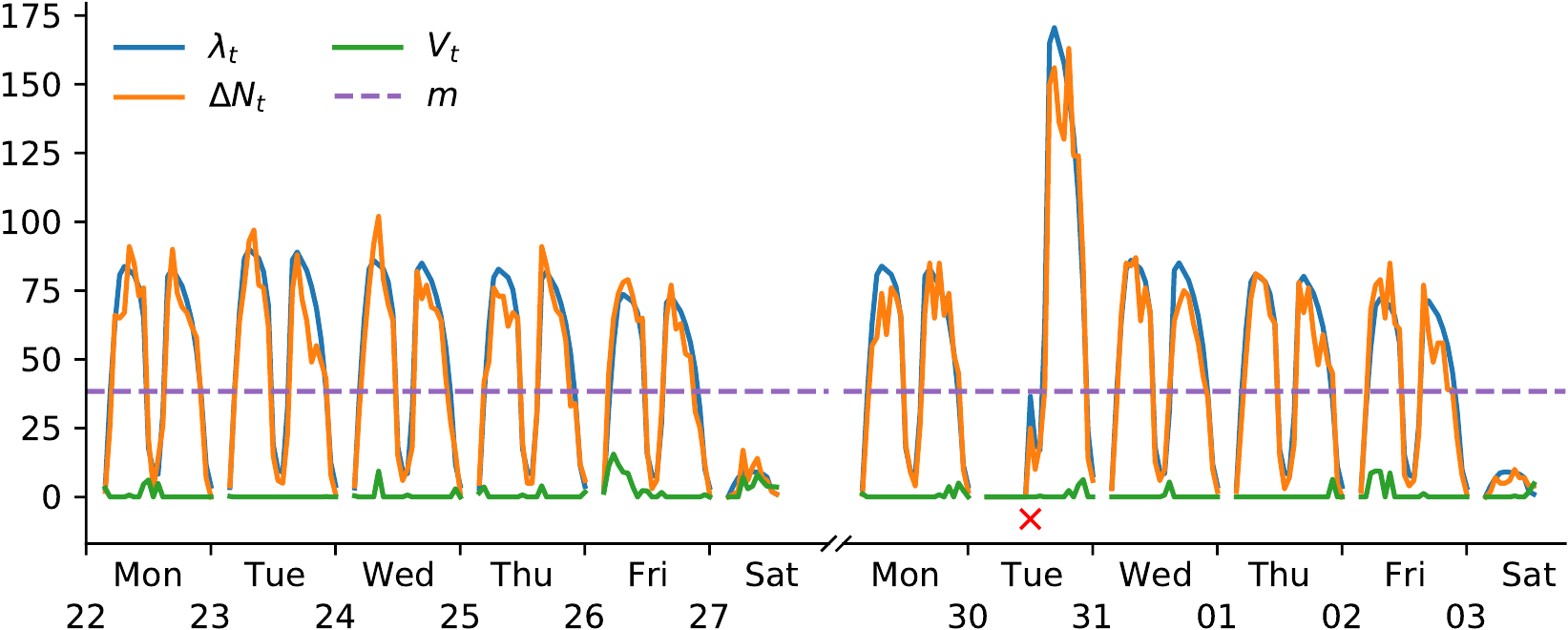}};
			\draw (-7, 2.5) node {(a)};
		\end{tikzpicture}
	\end{subfigure}
	\begin{subfigure}{\textwidth}
		\centering
		\vspace{-1em}
		\begin{tikzpicture}
			\draw (0, 0) node {\includegraphics[width=.8\textwidth]{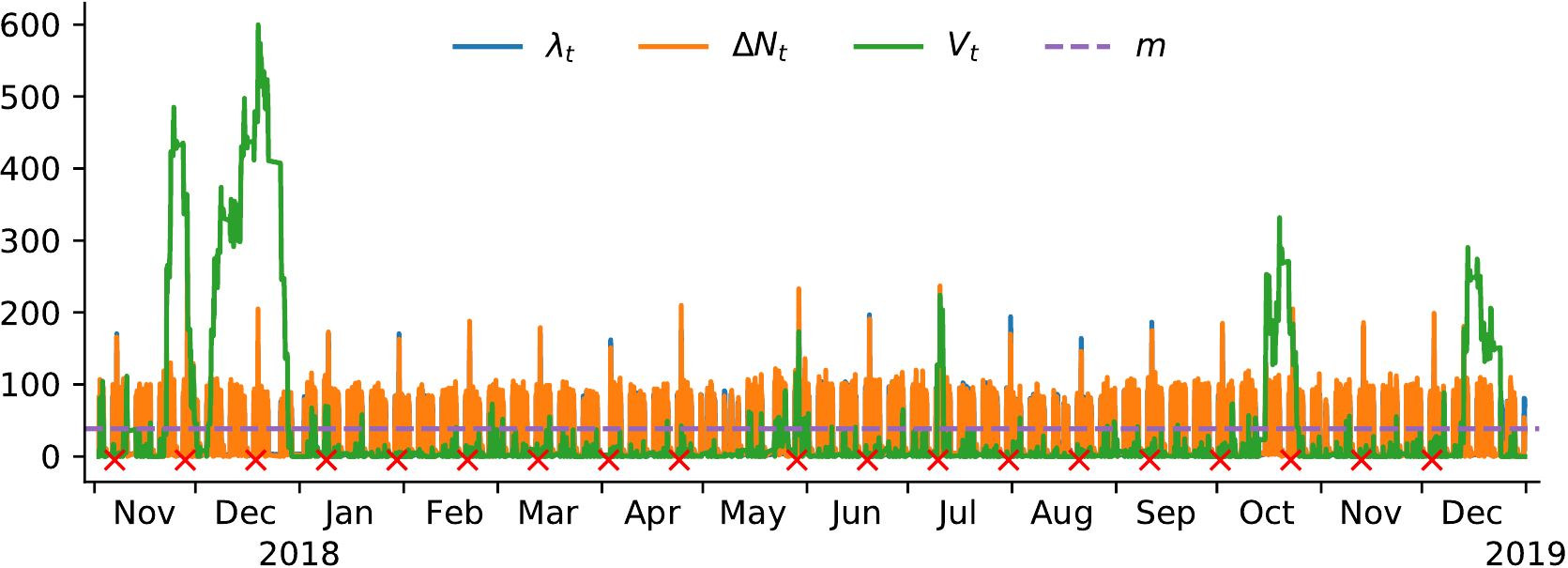}};
			\draw (-7, 2.5) node {(b)};
		\end{tikzpicture}
	\end{subfigure}
	\caption{The refitted CUSUM results over (a) a fortnight from 2018-01-22 and (b) the whole test set, where every third Tuesday (marked with a red cross) has had the morning calls shifted into the afternoon.}
	\label{fig:modified-tuesdays-fitted}
\end{figure}

However, the relevance of the CUSUM depends on the type of event: here, we use a very caricatural example. We could imagine that, if surges occurred regularly after a closure, this would have been incorporated in the model. In that case, the CUSUM handles very well this type of feature: in \autoref{fig:modified-tuesdays-fitted}, one can see that the CUSUM works well for this path-dependent intensity case. In particular, when nothing unusual happens except this expected surge on the Tuesday afternoons, the CUSUM remains close to zero. \\

Another possibility could be to run a double-sided CUSUM algorithm that aims at detecting an increase or a decrease in the intensity. In the same vein, some machine-learning type algorithms like NuPIC (Numenta Platform for Intelligent Computing) have been proposed to measure the deviation between the call numbers that are forecasted by the algorithm and the real outcomes \cite{ahmad2017unsupervised}. The NuPIC algorithm is based on a neuroscience-inspired type of neural network called a Hierarchical Temporal Memory (HTM), which aims to be a more realistic approximation of the brain than traditional artificial neural networks. The HTM is used to create a prediction of the next step ahead in a time series. NuPIC creates a prediction for each new data point as it arrives, and the distance between the predicted and observed values is used to determine if a change has occurred. All inputs and outputs to the HTM are represented as large sparse binary vectors, and so a distance metric to assess the accuracy of a prediction is based on the number of overlapping bits between the predicted value and the observed value. As each new data point arrives, a prediction is made, this overlapping binary distance which they call the `anomaly score' is used to determine the likelihood that this data point is anomalous, and finally they apply some smoothing to these anomaly scores to get something roughly equivalent to the $V_t$ in the CUSUM algorithm. \\

\begin{figure}[t]
	\centering
	\begin{subfigure}{\textwidth}
		\centering
		\begin{tikzpicture}
			\draw (0, 0) node {\includegraphics[width=.8\textwidth]{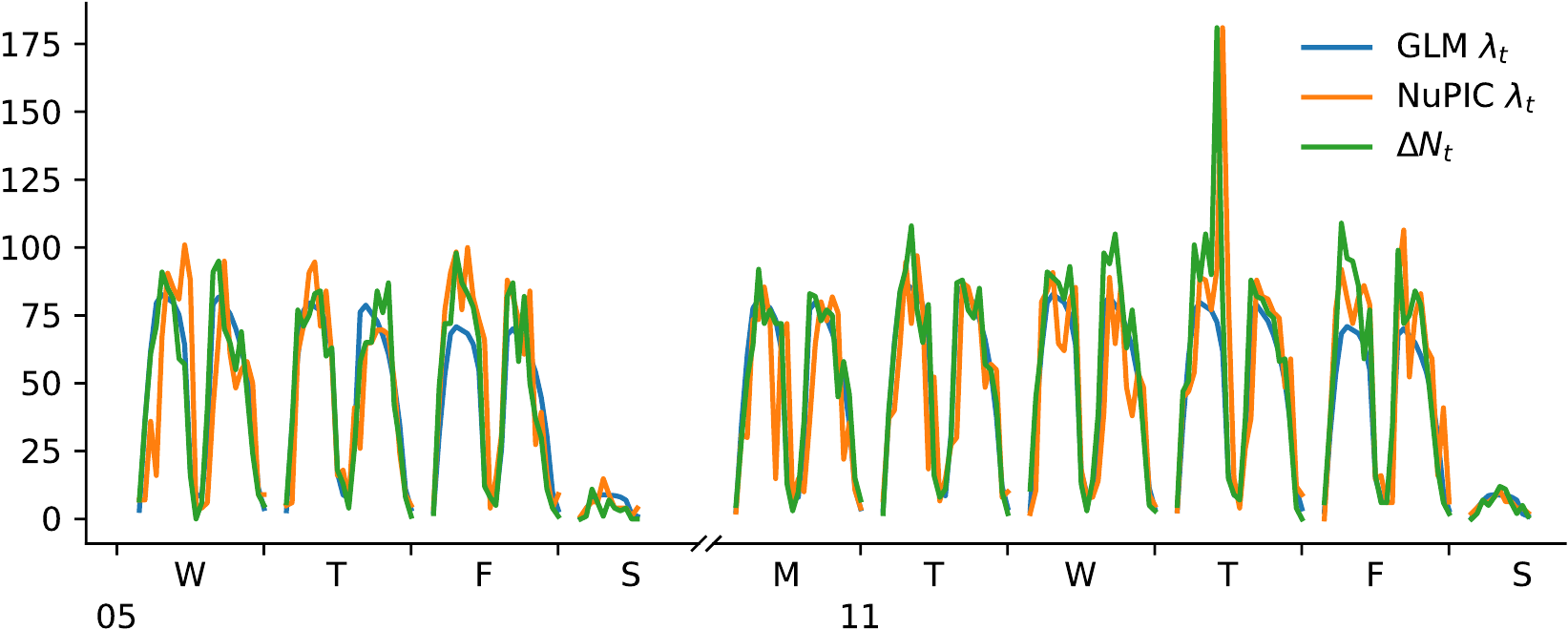}};
			\draw (-7, 2.5) node {(a)};
		\end{tikzpicture}
	\end{subfigure}
	\begin{subfigure}{\textwidth}
		\centering
		\vspace{-0.5em}
		\begin{tikzpicture}
			\draw (0, 0) node {\includegraphics[width=.8\textwidth]{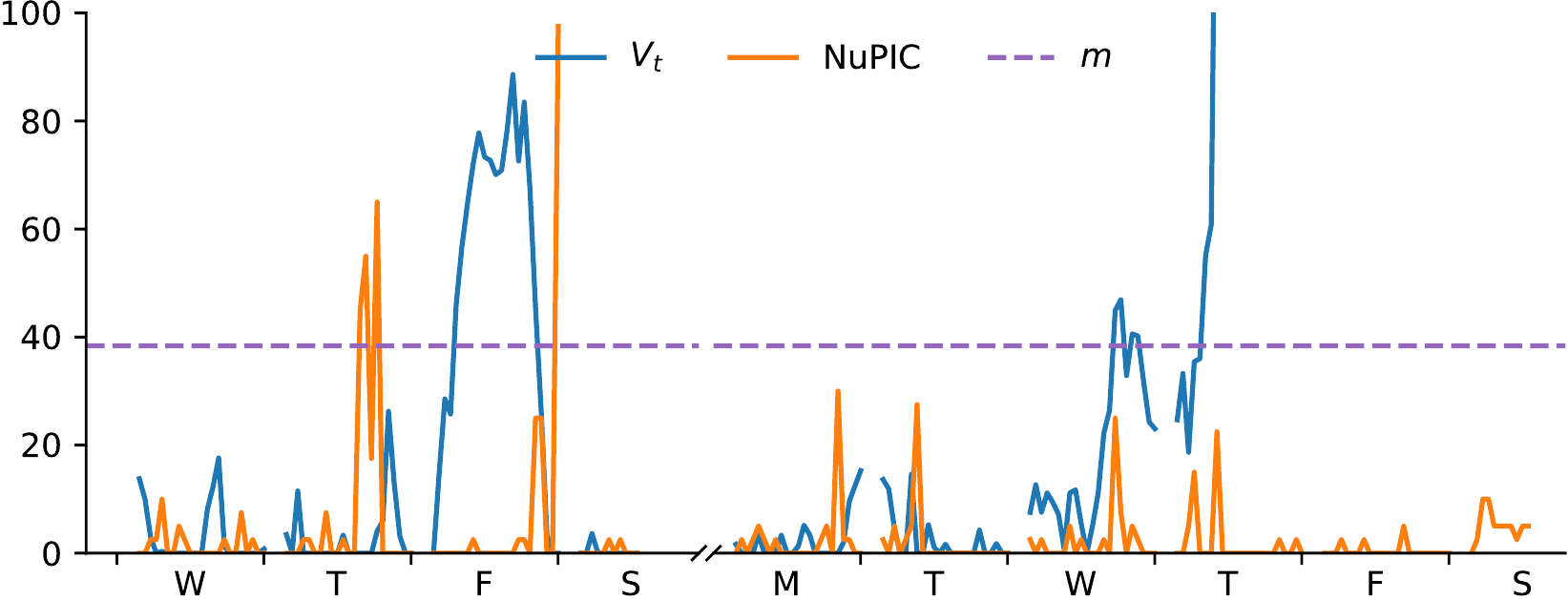}};
			\draw (-7, 2.5) node {(b)};
		\end{tikzpicture}
	\end{subfigure}
	\caption{Comparing the CUSUM and NuPIC results over some weeks from 2018-12-05. (a) The predicted number of calls for each algorithm and the observed number. (b) The level of anomalous behavior detected by each algorithm (NuPIC is scaled to take values in $[0,100]$).}
	\label{fig:nupic}
\end{figure}

The results of the unsmoothed implementation of this algorithm, are compared with the ones of the CUSUM in \autoref{fig:nupic}. The NuPIC algorithm does not perform too badly. It correctly identifies the busy period on the first Friday, yet it fails to react significantly for the huge spike the following Thursday. The reason we do not apply the smoothing described in the original paper is that the smoothed result just described nearly every data point as anomalous. Even unsmoothed, in our setting, it seems that NuPIC produces too many false alarms as detection is triggered by call numbers that are lower or larger than forecasted, while CUSUM is reflected at zero in case call numbers are much smaller than expected. These mixed results for detecting this class of anomaly inside has been remarked by others \cite{d2019experience}. Besides, there are advantages to using CUSUM over algorithms like NuPIC for risk management purposes. Firstly, CUSUM is much easier to implement and to explain. Secondly, CUSUM involves limited expert judgment while algorithms like NuPIC require to fine tune several parameters and a lot of expertise. Finally, there is some theory backing CUSUM and guaranteeing both its optimality in some cases and its robustness.

\section{Conclusion}
In this chapter, we have shown that the CUSUM algorithm can be adapted to the context of count data featuring seasonality. We explained how to use in practice, and we compared its performance with basic and machine-learning type competitors. The conclusion is that CUSUM represents a quite universal and robust tool for risk management, as long as one understand its limits.

% \bibliography{detection}
% \bibliographystyle{plain}

\end{document}